\documentclass[sigconf]{aamas} 

\usepackage{balance} 

\setcopyright{ifaamas}
\copyrightyear{2023}
\acmYear{2023}
\acmDOI{}
\acmPrice{}
\acmISBN{}

\def\DNA{\mbox{\it DNA}}

\title{Agent-Cells with $\DNA$ Programming: A Dynamic Decentralized System}

\author{Arash Vaezi}
\affiliation{
  \institution{Sharif University of Technology}
 \city{}
 \country{}
  }
\email{avaezi@sharif.edu}

\begin{abstract}
This paper introduces a new concept. We intend to give life to a software agent. A software agent is a computer program that acts on a user's behalf. We put a $\DNA$ inside the agent. $\DNA$ is a simple text, a whole roadmap of a network of agents or a system with details. A Dynamic Numerical Abstract of a multiagent system. It is also a reproductive part for an \emph{agent} that makes the agent not only take actions and decide independently but also reproduce coworkers. In fact, we look at an agent as a cell in the body of an alive creature.
The $\DNA$ may illustrate the agent's duties and communication protocols. 
By defining different $\DNA$ structures, one can establish new agents and different nets for different usages. We initiate such thinking as \emph{$\DNA$ programming}. This strategy leads to a new field of programming. This type of programming can help us manage large systems with various elements with an incredibly organized customizable structure. An agent can reproduce another agent. We put one or a few agents around a given network, and the agents will reproduce themselves till they can reach others and pervade the whole network. An agent's position or other environmental or geographical characteristics make it possible for an agent to know its active set of \emph{genes} on its $\DNA$. The active set of genes specifies its duties. A gene is a number a code that is a subset of a $\DNA$. There is a database that includes a list of functions s.t. each one is an implementation of what a \emph{gene} represents. 
The \emph{genome} (the set of all the genes) is the same for all the agents, but the environmental or geographical conditions make only a subset of genes active for a specific agent. To utilize a decentralized database, we may use a blockchain-based structure.

This design can adapt to a system that manages many static and dynamic networks. This network could be a distributed system, a decentralized system, a telecommunication network such as a 5G monitoring system, an IoT management system, or even an energy management system.
The final system is the combination of all the agents and the overlay net that connects the agents. We denote the final net as the \emph{body} of the system.

\end{abstract}

\keywords{Software Agents, $\DNA$ Programming, Genes, Dynamic Networks, Telecommunication Systems.}

\newcommand{\BibTeX}{\rm B\kern-.05em{\sc i\kern-.025em b}\kern-.08em\TeX}

\begin{document}


\pagestyle{fancy}
\fancyhead{}


\maketitle 


\section{Introduction}
Biological and nature-inspired algorithms were considered previously~\cite{article-natureinspired,cite:journal-natureinspired,e23070874}. 
Nature-inspired optimization algorithms (NIOAs) are defined as a group of algorithms that are inspired by natural phenomena, including biological systems and physical and chemical systems~(A. Baskaran. et al. \cite{1}). NIOAs lead to essential branches of artificial intelligence (AI). A large number of NIOAs have been proposed, such as genetic algorithm~(M. Chawla. et al. \cite{2}), particle swarm optimization algorithm~(A. Colorni. et al. \cite{3}), differential evolution algorithm~(I. Fister. et al. \cite{4}), artificial bee colony algorithm~(D. Karaboga et al.\cite{5}), ant colony optimization algorithm~(A. Colorni et al. \cite{6}),  firefly algorithm~(X.S. Yang \cite{cite:firefly}), and harmony search algorithm~(Zong Woo Geem et al. \cite{13}). NIOAs have been reviewed comprehensively~\cite{cite:review2,14,cite:review3}.

Meanwhile, we know that Agent-oriented Programming (AOP) is a kind of computer programming that focuses on agents and their behavior (for more details see~\cite{SHOHAM199351} presented by Yoav Shoham). AOP differs mainly from object-oriented programming in that it relies on externally specified agents instead of objects. The agents can be considered as abstractions of objects.

 Wooldridge and Jennings in~\cite{jennings1995applying} defined an agent as a hardware or software-based computer system with the following properties: (1) Autonomy: Agents operate without the direct intervention of humans or others and have a kind of control over their actions and internal state. (2) Social ability: Agents interact with other agents (and possibly humans) via an agent-communication language. (3) Reactivity: Agents perceive their environment and respond in a timely fashion to changes that occur in it. (4) Pro-activeness: Agents do not act in response to their environment. Nonetheless, they can exhibit goal-directed behavior by taking the initiative.

AIMA~\cite{brewka1996artificial} stands for Artificial Intelligence, a Modern Approach. Russell and Norvig, the authors of AIMA, define an agent as follows: \textit{ An agent is anything that can be viewed as perceiving its environment through sensors and acting upon that environment through effectors.} The AlMA definition relies heavily on what we take as the environment and on what acting and sensing imply.

\setlength{\textfloatsep}{5pt}
\setlength{\intextsep}{5pt}
\begin{figure}[thp]
\begin{center}
\includegraphics[scale=0.4]{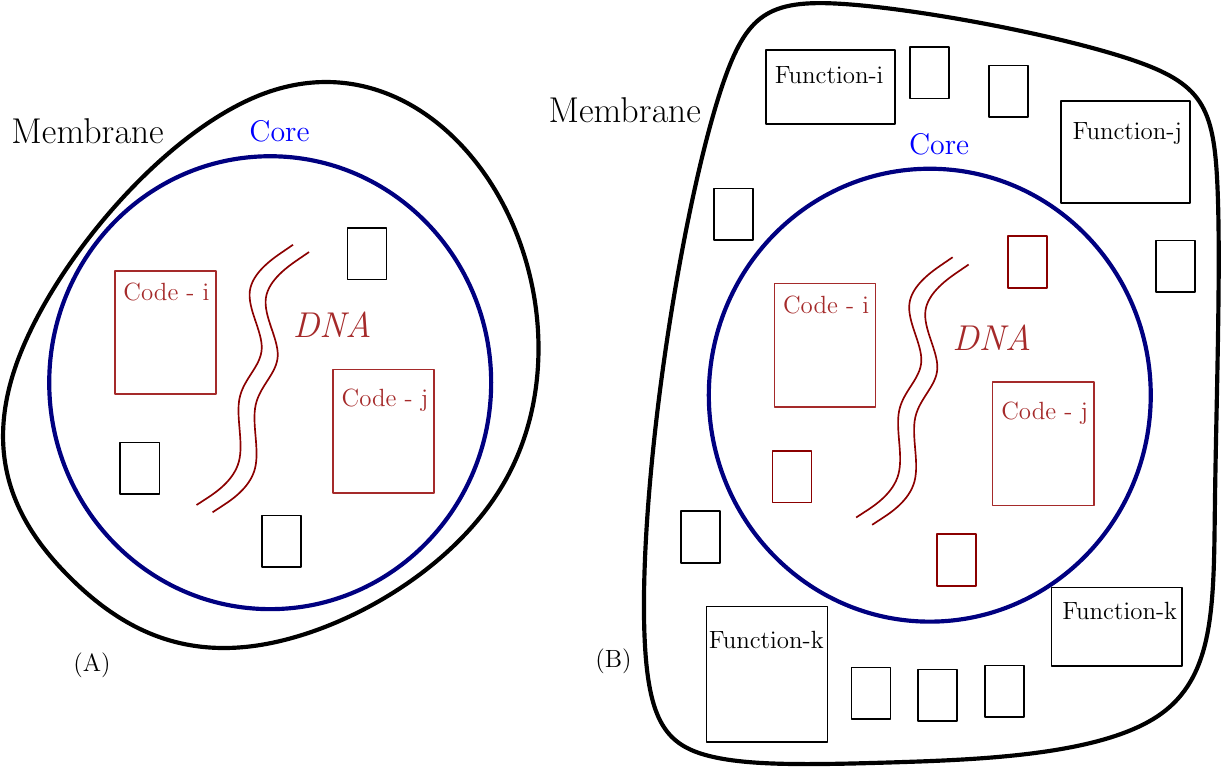}
\caption{(A) Illustrates a newly born agent. (B) Shows a filled membrane that contains the real implemented functions corresponding to the active genes of an agent after the reproduction process.}
\label{fig:cell}
\end{center}
\end{figure}  

In this paper, we define an agent to be a cell of the body of a network. The agent has a core, a membrane, and a $\DNA$ (see Figure~\ref{fig:cell}(A)). 
Note that our definition does not contradict the previous definitions of an agent. In reverse, the other systems can remarkably benefit from hiring the $\DNA$ capability inside an agent.
To the best of our knowledge, equipping a software agent with a $\DNA$ structure is initiated with this paper. This perspective was valuable when we tried to devise a strategy to move from a 4G telecommunication system to a 5G, where there are quite a few elements. Each element needs a different set of functional operations with tight deadlines. For example, an IoT device may have different requirements than a battery controller in an antenna. Decision-making and setting operational functions to be run for monitoring and controlling a device made us use software agents. The idea began when we wanted to provide an agent that could fork itself and spread throughout a network. Each agent was supposed to run a subset of a list of functions. This subset of functions may depend on a few characteristics, such as the agent's position in the given network. If we create an agent that contains the whole set of functions so that it can detect the position and run the corresponding subset of functions, there got to be a considerable code carried by every agent. This is not appropriate for coding and designing systems. We do not want to put a bunch of impractical codes in each network node.
 So, this idea comes to mind to see an agent as a living creature cell that carries a special data structure that carries only the required information denoted by $\DNA$. This structure consists of information demonstrated as a text. $\DNA$ has coding that specifies the functions an agent should do. The actual implementation of the required subset of functions is not at the core of the agent. Every agent, after being born, should load the related and required subset of functions from a shared database. Even the software architecture of an agent can be presented in its $\DNA$. The idea can be used to manage and monitor many distributed or decentralized systems. To provide a specification for a given network, we only need to design an appropriate $\DNA$ or create a new set of genes for a $\DNA$. 


The agents reproduce themselves and propagate to the whole given network. Finally, an overlaying net of agents is constructed, denoted by the \emph{body}. Based on the list of functions designated with the active genes in the $\DNA$ structure, the agents load their required corresponding functions to know their operations. The real implementation of the functions will be loaded into the agent's membrane (see Figure~\ref{fig:cell}).

 There could be various agents in each position of the network that works on different duties. An agent can decide and act accordingly; it can reproduce and make more co-workers. By programming a new $\DNA$, a new set of functions and structures are designed for the final overlay net of the agents (the body). So, an administrator only worries about the ways of programming a $\DNA$. 

All in all, we introduce a dynamic fractal concept, a new perspective.
Section~\ref{sec:agent} deals with the definition of an agent and a few characteristics of the agents. Section~\ref{sec:dna} \emph{suggests} a strategy for presenting a $\DNA$. Section~\ref{sec:example} contains a simple example of a network and a simple $\DNA$ programmed for the agents to reproduce in the proposed network, and Section~\ref{sec:specifications} explains a few characteristics the concept of agent cells with $\DNA$ can provide for a given network.
Live inventory detection is a crucial issue in the industry. Secion~\ref{sec:inventory} demonstrates how agent cells with $\DNA$ can help manage this problem. Of course, the new concept can be beneficial to solve several other issues that industries are dealing with nowadays. 

\subsection{The Background Infrastructure}
The idea presented in this paper is to use a data structure to develop a $\DNA$ for software agents to make a network in control and efficient management. We denote this idea by the agent cell with $\DNA$ protocol. In a dynamic network, a new list of elements may come and go to the previous network, and people need to work on a wide range of different software and drivers to make their system compatible with the new equipment.

To implement this protocol, initially, we need to come up with an infrastructure for the software installed on the elements of the given network. This infrastructure makes the software installed on the elements of a network support the agent cells with $\DNA$ protocol. The companies who create elements and their drivers and corresponding software should support the $\DNA$ protocol. We can use a library to be used by programming languages containing the procedures required for implementing this protocol.
This problem was previously encountered by scientists, too.
Java uses a middle language to make different hardware capable of running its codes. We can similarly bring up a new middle language for every software agent responsible for every element in the network; then, each software can support and understand the concept of $\DNA$. For example, considering telecommunication systems, designing a standard middle language can be beneficial to the companies that manufacture the telecommunication elements and their software and drivers. Another strategy is to use $C\#$ Open Source Managed Operating System (Cosmos), which is planned to make mobile agents. Agent libraries will also exist on Windows and other platforms to allow the agents to move between different operating systems.

Here, considering a given network with a number of nodes connected to each other, we are dealing with creating a new agent in a different node as a lightweight software agent that grows inside the new node. The creation process is initiated based on moving the required information (the $\DNA$) from another older agent. This is called reproduction. We denote each node of the network by an element. The word network itself is the topological graph connecting the elements. The body of the agents creates an overlay connection of agents on the underlying actual network.

It is not about controlling a new object remotely or deploying new objects. It is about creating a new agent cell. A new agent cell can grow differently from its father.  

Note that the approach we use to implement the agent cells with $\DNA$ protocol depends on the characteristics of the targeted network. Security might be a less challenging issue because the network is a private combination of elements and software. However, one might design a different approach to keep the protocol secure. We assume that the protocol is going to be used by a private network. So, suppose that the network nodes could access each other, and their software supports the reproduction process. Nonetheless, one may use a particular type of agent to create a security group of agents to keep the elements safe.

\subsection{$\DNA$ Programming}
By programming a new $\DNA$, one can support any dynamic changes to the network. 
Note that $\DNA$ is a text. So it is light and easy to move and copy. 
This point of view is helpful in upgrading the previous strategies, such as edge computing and many other previous methodologies. It can also make the network dynamically compatible with the new changes. We could not find any such point of view in previous works. This type of programming is new.
Incorporating $\DNA$-based programming technology into agent technology is an idea that suits programming certain types of agents dynamically. This technology can meet or adapt to the requirements that agent technology/multi-agent systems have established after many years of development.


The agents can grow through several elements, such as sensors and network edges. In addition, software or applications that monitor a component or element can be equipped with such a technology. Therefore, an agent with this capability can have a particular gene in its $\DNA$ for making decisions without human intervention. Or a gene that helps the agent train itself for a specific purpose. 

The agents equipped with $\DNA$ could enjoy having a part in their $\DNA$ that makes their communication clear. Any communication protocol can be used. We can enable them to access to several communication protocols, and based on their position, they can decide and choose an appropriate strategy.

There is a reproductive strand that specifies the way the agents should propagate. $\DNA$ owns genes that, based on the position of an agent in the given network, a subset of these genes is active. This subset makes the agent act accordingly and autonomously. Note that the actual implementation of the functions corresponding to the genes exists in a database. So, designers and administrators can easily update their functionalities, and the entire network can adapt itself to the changes rapidly.

If an update is required.
The corresponding gene of the modified procedures or drivers could be broadcast to the whole network for every agent to be able to update its $\DNA$. So, the agents are highly flexible and easily compatible with updates when equipped with $\DNA$.
Also, in the database, we can update the function corresponding to a gene and let all the agents use the updated function. To do this, we must set a procedure in the core of the agent to check for any updates periodically.

Each agent may learn from its environment as well, and the learning procedure can be turned on or off with the permission of the administrator who programmed the $\DNA$. So, one can make specific agents reactive, proactive, or behave any other behavior required for the target network. So, all in all, the $\DNA$ programming idea can make the agents or even software pieces of programs and applications or the elements of a decentralized system more intellectually capable and flexible. Note that this can also be a type of programming. That is when all the software elements of a network (including agents) are equipped with $\DNA$. Then, we only have to worry about programming the $\DNA$, and the whole body of the agents will react accordingly.


\section{Agent}
\label{sec:agent}
As we know, a cell is the basic functional unit of life. 
Here, agents are the cells of the body of the network. The agents are the functional units that can operate autonomously and without human intervention. 
This view provides a fractal system. A fractal system is an interactive system that can adapt to a changing environment. Such systems are characterized by the potential for self-recovery, self-organization, and even self-optimization. In a fractal system, semi-autonomous agents interact according to certain rules of interaction, evolving to maximize some measure like fitness or, here, the maintenance of the network. The agents are diverse in both form and capability, and they adapt by changing their rules and, hence, behavior as they gain experience. Neural networks and deep learning can help us develop more intelligent agents. Their adaptability can either be increased or decreased by the rules shaping their interaction. This can lead to the case that a fractal system has the potential for a great deal of creativity that was not programmed into them from the beginning. 

In the system introduced in this paper, we mention that one can define several \emph{types} of agents based on the set of functions that are required to be run in each node of a network while using this protocol. For example, a network that includes batteries may need a few functions to be set on an agent that is responsible for controlling a battery. So, we need an agent type for the batteries. Or, an agent type may deal with sending messages over the network. 

The required functions are stored in a database, and the $\DNA$ of an agent reveals what functions are required for a specific type of agent. 
An agent has two layers; Core, Membrane. The core contains the $\DNA$ and perhaps other more crucial and private information. The core is better to be implemented in a secure way. 

The membrane contains the codes of the required functions the agent has to run as its duties. In other words, the membrane contains the codes or information a newly born agent should receive from the database. The functions corresponding to the active genes will be loaded onto the membrane of the agent.

\subsection{Special Agents}
This section mentions two special agent types. These types are examples to show how flexible the agent cells with $\DNA$ protocol is.

Adverse environmental effects may cause the system to be encountered challenging, non-expected situations. Specific agents can be designed to handle such situations. Let denote such agents as \emph{emergency agent}. An agent which is encountered with such an effect calls an emergency agent to move to the location where an event occurred and handle the situation. Such a movement should be handled in the $\DNA$ of an emergency agent.

Any agent has a \emph{backup} twin, and the twin has no work to do; it just gets the information of the leading agent periodically. If something happens to the leading agent, the backup agent will work in the place of its twin agent. A flag in the $\DNA$ specifies which of the twins is enabled. There is a method in the core of an agent that sends a message to the twin agent if something happens to the agent and it cannot work correctly. The same message can call the emergency agents to inform admins about the situation. Emergency agents and backup agents made the overall system to be more reliable.  

To make the system secure, one may check the attacks discovered previously \cite{cite:review,li2017survey,vaezi:hal-03657061} and add appropriate types of agents to the system, or may use a reliable platform introduced previously \cite{citelor,cite:securesystems} for the body of the system.

\section{$\DNA$ Structure}
\label{sec:dna}
A $\DNA$ consists of numerical codes standing side by side on a list. Each code represents a function or a structure. 
Every agent cell has a $\DNA$.
The $\DNA$ of an agent specifies the functionality of that agent, its behavior, and connection rules.
In other words, a $\DNA$ structure is a list of digits.
A set of numbers in this list determines the operations that the corresponding agent should do. All such operations are implemented as a function. Accordingly, we have a set of implemented functions in a database, and every function has a unique ID. A function's ID is the same as the digits in a $\DNA$ structure. We denote the ID of a function as a \emph{gene}. There could be other additional structural and managing digits in a $\DNA$. 

A $\DNA$ of an alive creature encompasses biological information. It contains a sequence that encodes genetic information. Here, we use a list of digits as the operational information for an agent. 
We have several types of agents, and each one enables a unique specific subset of genes in the $\DNA$. By defining different genes in the $\DNA$ list, we can manage to have various different types of agents throughout a network. 

Different system designers may need completely different structures to be defined as $\DNA$s.
For a multilayer network equipped with several elements, each of which has requirements and operations with several tight deadlines, a well-designed $\DNA$ structure can assign specific agents to manage each piece of equipment individually and make real-time decisions faster. An example of one strand of a $\DNA$ structure is illustrated in 
Figure~\ref{fig:ourdna}.

A $\DNA$ structure may consist of two strands of codes. Each strand holds approximately the same information. The \emph{main strand} holds the information on the current status of the body, the neighbor agents, the agents' children, the agent's duties, the agent's type, and any other information that may be required for an agent in a specific network. For example, the agents should communicate with each other with a specific, well-defined, secure protocol, and without human guidance or intervention. This protocol is specified as a part of the leading strand of the $\DNA$. The main strand could be modified by upgrades or changes to the environment. The second strand is used for the reproductive system. It is useful to create a newly born agent cell with initial settings.
\setlength{\textfloatsep}{5pt}
\setlength{\intextsep}{5pt}
\begin{figure}[thp]
\begin{center}
\includegraphics[scale=0.43]{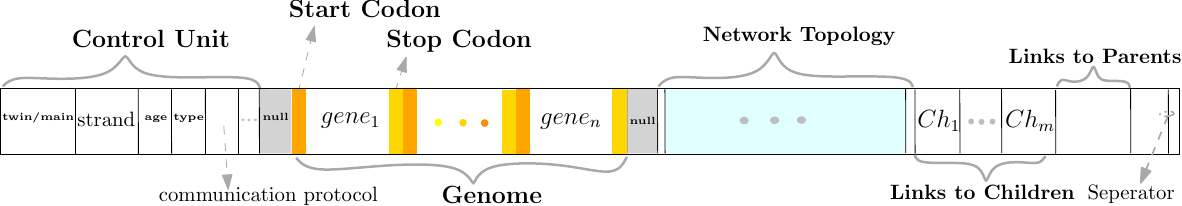}
\caption{This is an example of one strand of $\DNA$ of an agent cell.}
\label{fig:ourdna}
\end{center}
\end{figure}  

\subsection{Reproducing}
Recall that an agent cell has two main parts: a core and a membrane.  
Every agent has several procedures at its core. 
As mentioned before, $\DNA$ has two strands. The first strand is denoted as the \emph{main strand}, and the \emph{reproductive strand}, which contains the information that is required for making children for the agent.
For a fairly simple reproduction process, there are at least three primary procedures in the core of the agent cell: an initializer, a sequencer, and a $\DNA$ reproducer. 

When an agent cell decides to reproduce itself in another place.
The $\DNA$ reproducer in the source agent (the father) creates a copy of its main strand and produces the reproductive strand. This procedure sets the values in the reproductive strand as the initial standard values.
An initial instance of the agent cell with a copy of its core procedures will be created for the child. The $\DNA$ of the newly born agent owns only one strand which is the reproductive strand its father creates for it.
This is actually its main strand. The $\DNA$ reproducer for the child agent copies the values of the main strand of the child. The \emph{initializer} procedure reads the primary main strand and sets its values to be matched with the position and situation of the network element that holds this agent, the location of the child in the network topology, and the duties expected by the parent.

The \emph{sequencer} reads the $\DNA$ and based on the functional genes of the agent loads the corresponding implementation codes of the functions into the membrane of the agent. Note that the newly born child owns a core and an empty membrane.


The code of the implemented functions coming from the database formed the membrane of the child agent. So, the new child gets ready for acting. 

\paragraph{Dominant Gene Table}
There could be such positions in the given network where two or more agent cells try to reproduce a child into one specific element of the network. 
If there is more than one father for a child, we could deal with different strategies. We can either create different children inside one element and let them work side by side, or we can merge them into one child. The latter one seems to be more interesting. By programming a $\DNA$ for a network we define a table for genes. This table is called the dominant gene table (DGT). When two or more agent cells try to reproduce into one element, if there is already an agent cell in that position, the initializer detects this situation and makes all the required genes active. However, if there are non-identical sorts of genes the initializer activates only the dominant genes. So, the dominant gene table is beneficial for the initializer.

\textbf{Type of an Agent:}
As mentioned previously, every agent has a type. 
Different agent types use different subsets of genes in $\DNA$
The type of an agent may be determined via a code ID in an entry in the main strand of the $\DNA$ of an agent.
The reproductive strand may also contain the type of children of an agent. The initializer can read this part of the reproductive strand and decide the type of child for the agent.

\textbf{Age of an Agent:}
There is a specific entry in the $\DNA$ of an agent, which specifies the \emph{age} of an agent. The age is a positive integer. Each run of a function in the membrane of the agent increases the age by one. So, the agents who work hard get old sooner. So, after a specific period of time, we can detect the more active parts of the network, the bottlenecks, etc.

\section{System Specifications}
\label{sec:specifications}
Different programming of the $\DNA$ strands provides a different set of distinctive characteristics for the final body of the system. So, the system is quite flexible. This section mentions a few characteristics that an administrator may need to provide for a network.

\subsection{Scalability}
\label{subsec:scalability}
The reproductive system makes the proposed system highly scalable. This is, in fact, a highly under-control automated, scalable system whose scale does not affect its quality. 
Scalability is one of the biggest benefits of distributed systems. It is not just about adding more elements and resources. It is about controlling a big system as well as a small one. The agent cells can reproduce and monitor the system autonomously. To control a large-scale system with tight deadlines, the problem gets even more challenging. In the proposed protocol here, every agent can use an artificial intelligence built-in subsystem to get itself compatible with the environment and learn from the environmental conditions. 
Combining this system with a reliable previous management system such as blockchain or any other trustful system \cite{citelor} makes it secure on a large scale.
\subsection{Self-Monitoring}
\label{subsec:monitoring}
We can add a particular type of agent that is like a token and moves around the body of the overlay network to monitor the system. To have a more secure system, we may let some monitoring agents move randomly and check the behavior of other agents. 
An agent-based system whose agents are equipped with AI and $\DNA$ has, in fact, a built-in management and monitoring system. Considering to each element of a network, there could be an exclusive agent that watches the behavior of that element. The correct behavior of an element can be defined as rules. There are equivalent functions corresponding to the operations of the element. These functions are stored in the database. The genes in the $\DNA$ specify the functions of the agent. So, the agent loads the functions from the database and monitors the element's behavior.  

\subsection{Adaptability}
\label{subsec:adaptability}
For a system to be adaptable, it should detect the environment and its position, the requirements and its level of power, a few rules, and its security level. So, for a system that works based on the agent cells, perhaps a class of agents is required in every node of the system. Denote such a class as a cluster. Each cluster can have a shared memory and inner communication protocol. The network that contains the clusters as nodes may use another outer communication approach. This system defines a set of genes in $\DNA$ for the agents inside the cluster and another set of genes in $\DNA$ for agents which monitor the outside network. In such a system, corresponding to each specific element, there is a cluster that checks the element and its environment and adapts itself to the situation of that element, and the cluster has the power to change itself with the changes of the network around that element. Controlling, monitoring and managing are among the duties the agents join in a cluster have in their set of functional operations. 
\subsection{Resource Management}
\label{subsec:ResourceManaging}
Agent-Cells could be incredibly useful in managing and monitoring several resources in a large-scale system. An example of such a system is an energy management system. 

To improve the energy efficiency of a whole network, particular agents may be assigned to control the performance or energy consumption of every network's machinery, device, or component. Observe that every device has the best state and performs well if we set it in that state. So, the element can be monitored and managed by particular agents to find such an efficient state(s).
In fact, an agent can check the status of a particular component and learn how to set the controlling parameters of that component. The age of such an agent specifies if it is a well-learned agent. So, we can train various agents by monitoring and controlling every component of an energy management system. The final body of such a network of agents brings an incredible energy management system. Of course, by updating the $\DNA$ of such a network, the system can support managing a dynamic energy management system. Therefore, $\DNA$ Programming can become an intelligent system that offers a very high level of self-adaptability and flexibility.

Let's see an example. In 20201, Arcos-Aviles et al.~\cite{arcos2021energy} suggested a Fuzzy Logic Control (FLC) based Energy Management System (EMS), smoothing the grid power profile of a grid-connected electro-thermal microgrid. This system consists of three blocks, where the first one uses an FLC component to compute the preliminary value of the grid power profile due to the error in the microgrid (MG) power balance forecast and the battery state of charge. For more details, please refer to the original paper. Arcos-Aviles et al. consider their approach a low computational complexity due to the simplicity of the offline trained fuzzy controllers that can be embedded in low-cost digital platforms. However, a drawback of the system above is the availability of historical data for a year, mainly on the consumption side~\cite{arcos2021energy}. The $\DNA$ Programming system can interestingly offer improvements from different aspects. Initially, an agent cell with a $\DNA$ can receive all functionalities required for the above EMS system. Observe that the deep learning module is already on the consumption side. Thus, it automatically solves the issue of the data remaining there for a year. Also, when the agent updates its age, it becomes an expert agent. Accordingly, the agent can predict the behaviors of the EMS more accurately, which suggests an overall improvement in the quality of managing the EMS mentioned above. In addition, since the $\DNA$ Programming is a fractal system, the agent recovers automatically, in case needed. Thus, the availability of the above EMS would improve as well. Even more, if a change in the policies of the EMS occurs, the agent cell conveniently adapts it.

\section{An Example }
\label{sec:example}
\begin{figure}[thp]
\begin{center}
\includegraphics[scale=0.41]{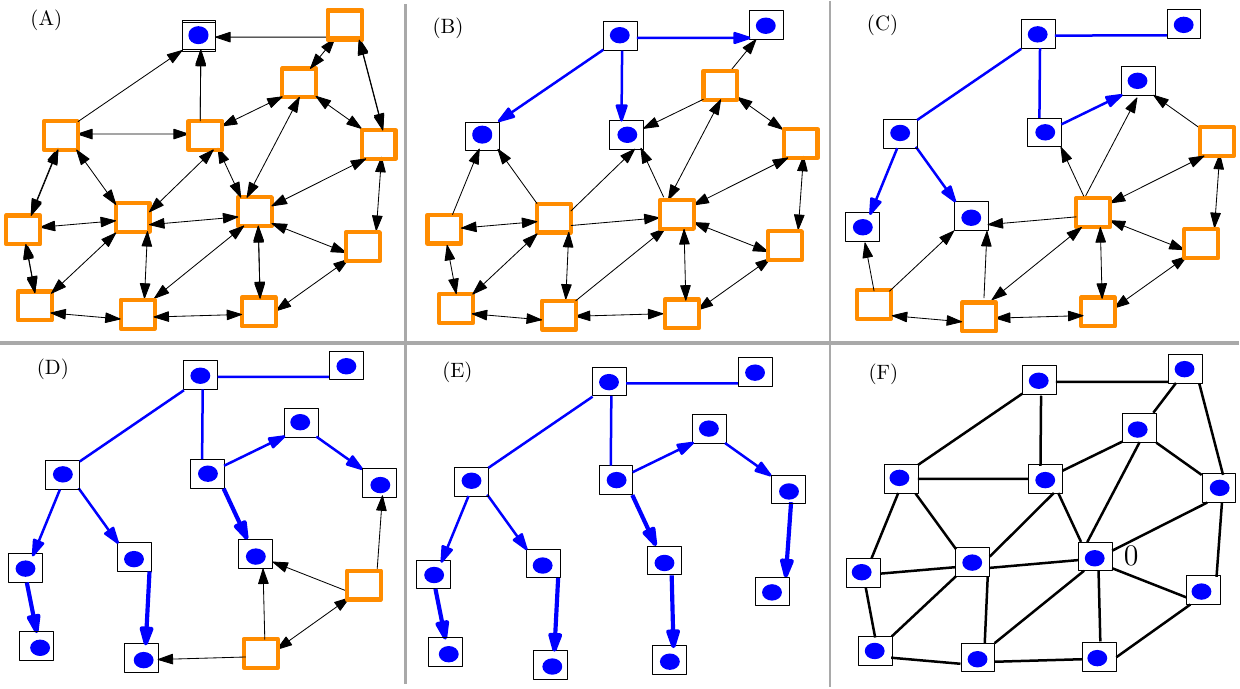}
\caption{Network topology pervaded by agent cells.}
\label{fig:prevading}
\end{center}
\end{figure}

This section illustrates an example to show how we can create a body of agent cells with $\DNA$. Suppose we are given an input network with an initial topology. We suppose that every network element supports the presented protocol in this paper, i.e., agent cells with $\DNA$. That is, every element seeks agents with $\DNA$. Figure~\ref{fig:prevading} illustrates an example of a given network. If we fertilize a random element of the given network with an agent cell, the network gets filled with the agent cells. 

To pervade the network, agents should reproduce through all the elements. Every element not equipped with an agent sends messages requesting an agent cell. An agent inside an element receiving messages reproduces a newly born agent cell for the requested element. 

For the reproduction process, consider three procedures in the core of the agent cell: an initializer, a sequencer, and a $\DNA$ reproducer.

The $\DNA$ reproducer copies the primary strand of the $\DNA$ in the very moment that a newly born agent is created. For reproduction, the $\DNA$ reproducer in the source agent creates a reproductive strand and a newly born agent. 
This newly born agent will be moved to the element that handles the child.

The newly born agent has only one strand, which is its main strand. The $\DNA$ reproducer copies this strand. The \emph{initializer} procedure reads the main strand of the $\DNA$ and sets its values to be matched with the position and situation of the network element that holds this agent cell. The \emph{sequencer} reads the genome and finds active genes of the agent cell. Based on the active genes, the sequencer loads the corresponding implementation codes of the functions and procedures into the membrane of the agent. 

So, note that during the reproduction from a sourcing agent to a child, the reproduction strand does not have all the values set precisely. 
The values of the reproduction strand will be completed by the initializer when the strand is set as a main strand in the core of the child cell.


\setlength{\textfloatsep}{5pt}
\setlength{\intextsep}{5pt}
\begin{figure}[thp]
\begin{center}
\includegraphics[scale=0.43]{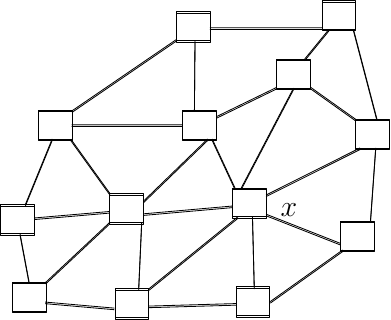}
\caption{An example of the topology of a network.}
\label{fig:basicskeleton}
\end{center}
\end{figure}

Figure~\ref{fig:basicskeleton} illustrates the given network inventory of the example. 
The network elements will be equipped with agent cells as illustrated in Figure~\ref{fig:prevading}. In the initial step, a network element is equipped with an agent cell. All other agents are seeking an agent cell. The initially equipped element fertilizes three other elements in the next step illustrated in Figure~\ref{fig:prevading}(B). The already-equipped elements with agent cells do not send messages to their neighbors. In this example, the network element that receives a newly born child from more than one agent cell should eliminate all except one.
In this example, all the elements of the final network should be equipped with agent cells successfully, i.e., each element should own exactly one agent cell. 

\setlength{\textfloatsep}{5pt}
\setlength{\intextsep}{5pt}

For this example, suppose an agent can send messages to all its neighbors. So, it does not matter who is the father of an agent cell. Since it does not matter which agent cell will be the father of a child, an element can use any strategy to eliminate all answers from neighboring agent cells except one. Figure~\ref{fig:prevading}(D) and (E) illustrate how this network is pervaded with agent cells. Ultimately, all elements are equipped with agent cells with $\DNA$. The sequencers will read the main strand of the $\DNA$, and the right subset of functions will be loaded to the membrane of each agent in the memory of the element.

\setlength{\textfloatsep}{5pt}
\setlength{\intextsep}{5pt}
\begin{figure}[thp]
\begin{center}
\includegraphics[scale=0.43]{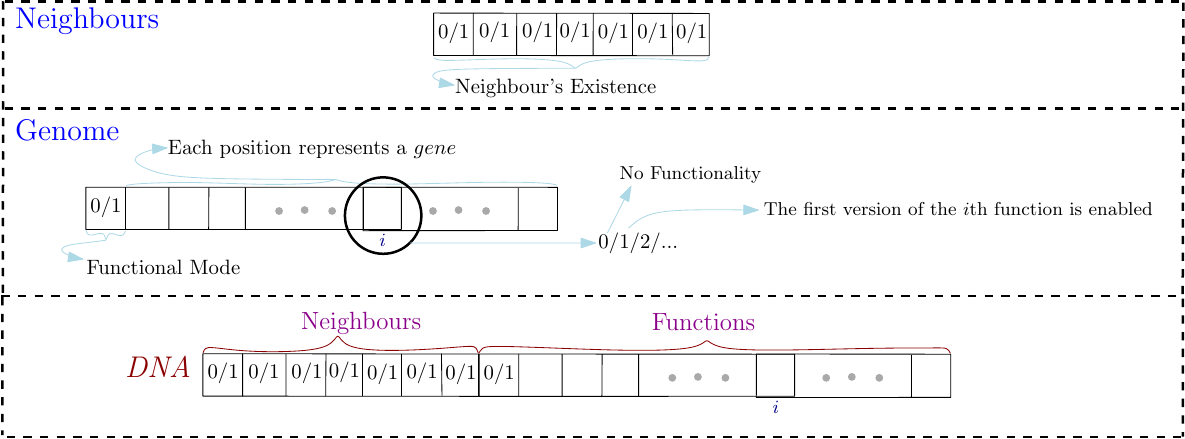}
\caption{The part of a programmed $\DNA$.}
\label{fig:ourdna}
\end{center}
\end{figure}  

Figure~\ref{fig:ourdna} illustrates a simple $\DNA$ strand for the above-mentioned network. There are two pieces.: Neighbours: This part demonstrates the availability of the neighbors of the element. A maximum of seven neighbors are allowed for an agent to be connected with. This part store a local topological structure of the network for the agent.

The second part is denoted as the "Genome". Each cell of this genome illustrates a gene. Here, a gene is an integer number larger or equal to zero. If a gene is zero, the corresponding function is disabled for the agent. Otherwise, the integer indicates the version of the corresponding function to that gene that should be uploaded and could be executed by the agent cell.

\setlength{\textfloatsep}{5pt}
\setlength{\intextsep}{5pt}
\begin{figure}[thp]
\begin{center}
\includegraphics[scale=0.5]{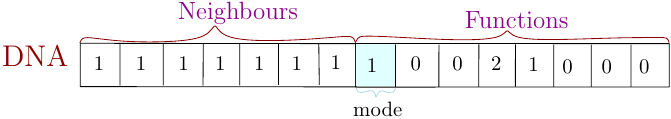}
\caption{The main strand of the $\DNA$ of node $0$.}
\label{fig:sampledna}
\end{center}
\end{figure}  
See Figure~\ref{fig:sampledna}. This is the main strand of the $\DNA$ of the element specified by the number $0$. This node has $7$ neighbors. 
The node is functional, and there are two active genes out of seven genes in the genome of this agent. The third and fourth genes are active. The second version of the procedure corresponding to the third gene and the procedure corresponding to the first version of the fourth gene will be loaded into the membrane of the agent cell that exists in the element $0$.

Now that we have come up with the structure of the $\DNA$, we need to see the requirements of the network and create the list of the functions. Note that in practice, the requirements of a network are already known. The functions, the procedures, the drivers, and all the actions of a system are clear to the managers of a system themselves.
\subsection{A Senario}
Consider a very simple and elementary sensor network with the initial inventory illustrated in Figure~\ref{fig:skeleton}. A sensor network comprises a group of small, powered devices and a wireless or wired networked infrastructure.
This scenario illustrates a simple picture of the body of agent cells on a given network. Consider the following assumptions:
\begin{enumerate}
    \item 
In decentralized systems, it is important to have a coordinator that can make the decisions and broadcast important information to the whole network. Here, we set the node with the most number of neighbors to play the role of a coordinator. If something happens to this coordinator, it is required to pick another agent with the most number of neighbors less than the actual previous coordinator. So there is a backup, but only one backup in this example. See
Figure~\ref{fig:skeleton}, node $0$ should be assigned as the coordinator. There is another element that has six neighbors. The agent cell placed on this element should be set as the alternative coordinator.

\item 
Suppose every element of the network owns a battery. Each agent cell that lives on an element has an age. The older the agent cell is, the less battery is available for the corresponding element. 
Suppose when the age gets close to a number, it realizes that the battery is low enough, and the agent needs to run a procedure that corresponds to a gene. And also, the agent needs to generate an alarm. The agent reports this alarm to the coordinator by informing its neighbors. The neighbors should send this type of information to their neighbors till it reaches the coordinator. Then, the coordinator decides what to do. 

For example, when an agent realizes the battery is too low, it may be allowed to connect all of its neighbors and moves out of the network.

\item 
In this example, every agent periodically sends acknowledgments to its neighbors. So, after a specific number of consecutive periods, say three, if a neighbor does not return an acknowledgment, we can assume that the element is unavailable. The agent that distinguishes the lack of an element sends the information to its available neighbors till this information reaches the coordinator. The coordinator reports this information in a database. If an element needs to be added/eliminated, the coordinator knows about the change. Adding a new element is easy because the new element sends requests for reproducing an agent cell to its neighbors, and a newly born agent will be reproduced for the element. The eliminating procedure, however, is based on the rules that are specified by the functions and procedures corresponding to the genes.
\item 
Set the agents around the coordinator as first-class agents. These agents send the information to the coordinator easier. Other agents are normal agents. The agent types have different duties. For example, suppose the first-class agents wait five periods for their neighbors to respond.

\item 
Assume that all the agents are functional. If something happens to an element, the agent's status will change to non-functional. Other agents cannot receive any acknowledgment from a non-functional agent, so they send reports to the coordinator. The element should be eliminated, or the agent should be reborn with a reproduction process again.
\end{enumerate}

List of Genes:
\begin{enumerate}
    \item F[1]: 
  This gene enables first-class agents to communicate faster.
If something happens to the coordinator, these agents should react accordingly. They know the address of the backup coordinator, and they can choose the new coordinator immediately.

\item F[2]: 
    This gene is active for agents with an even number of neighbors. If an agent has two neighbors, it should send a request to the coordinator to introduce one more neighbor. If an agent has four neighbors, it can eliminate one of its neighbors to save more energy. If the agent has six neighbors, it is the backup of the coordinator. So, it should keep its neighbors, and if something happens to the coordinator, then this agent should send a request to one of its neighbors and get connected with another agent. 
    
    So, the minimum number of neighbors is three. And also, the coordinator always has seven neighbors.

 \item F[3]:
 Version 1 of the procedure corresponds to this gene checks if something happens to the element or its corresponding agent cell and creates an alarm. Then, the alarm will be sent to the neighbors till it reaches the coordinator. 
 
 Version 2 is written for the coordinator itself. In this version, the coordinator sends a message to the node with six neighbors to select the next coordinator. The coordinator should be changed till the main coordinator gets ready and healthy again.
    
\item F[4]: 
This gene enables the procedures related to the coordinator. Decision making, communicating among the agents. Reporting the alarms and all the duties related to a coordinator will be enabled for an agent with this gene active in its main strand of the $\DNA$. If we set the coordinator to another agent, this gene must get activated for the new coordinator. Version 2 of this gene enables the backup coordinator to reach the corresponding procedures if something happens to the coordinator.

 \item F[5]:
 Consider an agent $A$ and one of its neighbors $B$.
 This gene enables $A$ to get connected with a neighbor of $B$. This gene is active when $A$ and $B$  both have three or fewer neighbours.

 \item F[6]:
 This gene enables an agent to remove one of its neighbors. Every agent should have at least two neighbors, and there should be a connection between every agent and the coordinator. This function is proper when the agent does not have much energy available and needs to go to a power save mode. The gene is active for those agents with four or five neighbors.

 \item F[7]:
  This gene also enables the agent to change the structure of the overlay network (the body). The gene links the neighbors of the agent to another agent. Then, it can get eliminated. Namely, if the element does not work anymore, if the corresponding agent works correctly, it can eliminate the element from the network.
Version 1 of this gene lets the agent connect all of its neighbors to each other and eliminate itself. This version of the gene should not be activated for those agents that have more than three neighbors.

Version 2 of this gene connects two of its neighbors. This is useful when the neighbors try to communicate with each other with lots of information passing through the middle node. This version is active for the agents with four neighbors.

\end{enumerate}

\paragraph{The Initializer}
The initializer procedure in every element checks the condition of the element and sets the values of the main strand of the $\DNA$ in the corresponding agent cell. 

See Figure~\ref{fig:skeleton1}. There are thirteen elements and their corresponding agent cells in the body of the network. Node numbers indicate the agent cells live on each element. Node Zero is the coordinator that has seven neighbors.  

Nodes indicated by $4$, $5$, $6$, $8$, $9$, $11$, and $12$ include the first calss agents.  Node $8$ has six neighbors, so it is the backup coordinator. 

\setlength{\textfloatsep}{5pt}
\setlength{\intextsep}{5pt}
\begin{figure}[thp]
\begin{center}
\includegraphics[scale=0.43]{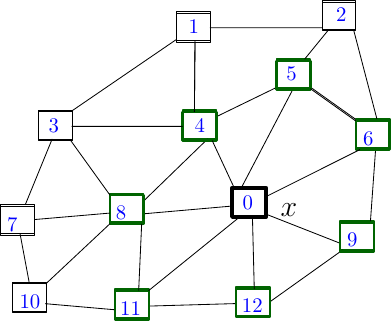}
\caption{An example of the topology of a network.}
\label{fig:skeleton1}
\end{center}
\end{figure}  

All in all, the initializer makes the initial main strand of the agents in this network, as it is illustrated in Figure~\ref{fig:dnas}. 

\setlength{\textfloatsep}{5pt}
\setlength{\intextsep}{5pt}
\begin{figure}[thp]
\begin{center}
\includegraphics[scale=0.43]{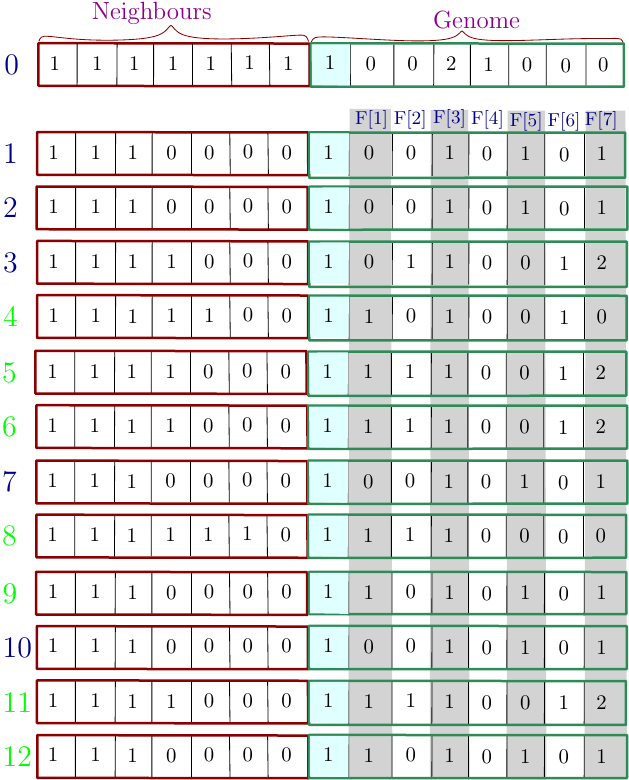}
\caption{The initial $\DNA$ of the network of the example.}
\label{fig:skeleton1}
\end{center}
\end{figure}  

Now, with the genes and their corresponding functions, the network inventory keeps being stable by the rules. The network could be managed automatically. 

\paragraph{A Managing Senario}
Consider a low battery alarm on node $4$. The gene $F(3)$ makes the agent send the situation to the coordinator. Based on gene $F(6)$, the agent needs to get disconnected from one of its neighbors, and the eliminated connections should be between nodes $4$ and either $5$ or $3$. That is because node $1$ has only three neighbors, and node $0$ is the coordinator that cannot have fewer neighbors. Also, node $8$ is the backup of the coordinator that cannot have fewer neighbors. Between node $5$ and $3$, node $5$ is a first-class node. So, the connection between $4$ and $3$ should be eliminated. This happens by sending number three to node $4$ by the coordinator to inform that node $4$ could run gene $F(6)$ by parameter three.  

Node $3$ detects the elimination of node $4$ and, based on gene $F(3)$, sends a message to the coordinator. The gene $F(2)$ eliminated node $4$ from the neighbors' list of node $3$. This gene also changes the $\DNA$ to deactivate gene $F(2)$ because node $3$ has three neighbors now.
\section{Live Inventory Detection}
\label{sec:inventory}
Live inventory detection is among the fundamental problems in the industry in IoT, Telecommunication, and many other networks. We intend to manage asset inventory efficiently with powerful automation using agent cells with $\DNA$. There are many challenges when keeping track of data inventory, network, wireless and mobility assets. So, with the right inventory management system and automated network maintenance, one can eliminate many complicated challenges and assign an automated system to manage them.

Several works tried various methodologies for this problem; object recognition and localisation~\cite{hebbalaguppe2017telecom}, deep learning methodologies~\cite{deng2021deep}, using tags and tracking appraches~\cite{TEJESH20183817}, and cloud computing~\cite{bose2022design}. However, the live essence of the inventory makes the problem harder. An automatic system has to deal with quite a few issues.

The elements of a live inventory are dynamically moving in and out. So, we have to monitor a given network dynamically and manage to install various drivers and software procedures to control the new units. However, using agent cells with $\DNA$, the new drivers and procedures will be stored in the database, and we know that the new element will be caught by the agent cells via the reproduction system. The new functions are available via new genes or new versions of the previous genes. The initializer that sets the values of the $\DNA$ sets the genes for the main strand $\DNA$ of the new element. So, the $\DNA$ of the gene contains new genes, and it is compatible with the network.

Consider the body of a cold-blooded creature. These animals have amazing adaptations that help them live even if they lose a part of their body. They can even recreate a lost part (organ) of their body. Here, we created a body of agent cells covering all through a given network. 
In each locality of this body of agents, they can check around and see if something happened to the previous status of the inventory. Accordingly, all change events in the network can be detected fast, and the recovery may start immediately and even automatically. In such a network, the $\DNA$ of an agent is programmed to let it check its neighbor agents in a specific periodic manner. Thus, the changes in the topology of the network get detected as expected. Moreover, emergencies or unexpected events will be recognized in a reasonable time with the smart agents. The dynamic $\DNA$ structure of the agents makes it possible to make the agents act differently in confronting different situations. 

\section{Discussion}
\label{sec:discussion}
We introduced a platform where there is an agent with a $\DNA$ structure. This agent reproduces itself through a network. Consequently, the crowd of agents will spread through the network. The agents can decide individually or via a voting mechanism with each other to manage a specific part or element. These agents may be equipped with an AI system or any advanced programming strategy. 

The idea uses a biological point of view, a new perspective, and another nature-inspired approach. The definition presented here is a new capability for a software agent compatible with all the previous. This capability makes the existing agents even more powerful and adaptable and could lead to a new type of programming. Namely, by adding a new package or library, one can make software products and agents compatible with the concept of $\DNA$. Then such a software or agent would have the potential to be programmed by programming the $\DNA$ structure. So just changing the $\DNA$ makes the agents and their behavior to be different. 

One can create an entirely different system by programming a different $\DNA$. Agents can decide to program their children to have a different $\DNA$. The first $\DNA$ for the very first agent in a network can change the whole behavior of the network. This is the idea that we initiate and denote as $\DNA$ programming.

We will empower this point of view in the future with more creative algorithms and strategies that can help agents act smarter and quicker in a dynamic network.
This research introduces the fundamental basics of $\DNA$ programming. The exact implementations and applications can be described accordingly in collaboration with the industry. 

Previously, based on the needs of a business, an application was required. And, to make it functional, designing new hardware was inevitable.
Later in time, a universal computing machine could support many applications by getting programmed differently, and programming came into the world. In the future, the $\DNA$ programming might offer to modify a $\DNA$ only and produce several systems instead of programming and designing a whole software the way we see it nowadays.


\enlargethispage{\baselineskip}



\appendix



\bibliographystyle{ACM-Reference-Format} 
\bibliography{sample}


\end{document}